\documentclass{pnastwo}

\url{www.pnas.org/cgi/doi/}
\copyrightyear{2008}
\issuedate{Issue Date}
\volume{Volume}
\issuenumber{Issue Number}
\begin{document}

\title{Strain and the optoelectronic properties of non-planar phosphorene monolayers}

\author{Mehrshad Mehboudi\affil{1}{Department of Physics. University of Arkansas. Fayetteville AR, 72701. USA},
Kainen Utt\affil{1}{Department of Physics. University of Arkansas. Fayetteville AR, 72701. USA},
Humberto Terrones\affil{2}{Department of Physics, Applied Physics, and Astronomy. Rensselaer Polytechnic Institute. Troy, NY 12180. USA},
Edmund O. Harriss\affil{3}{Department of Mathematical Sciences. University of Arkansas. Fayetteville AR, 72701. USA},
Alejandro A. Pacheco SanJuan\affil{4}{Departamento de Ingenier{\'\i}a Mec{\'a}nica. Universidad del Norte. Barranquilla. Colombia}
\and
Salvador Barraza-Lopez\affil{1}{}}

\contributor{Submitted to Proceedings of the National Academy of Sciences of the United States of America}

\significancetext{Phosphorene is a new 2-D atomic material, and we document a drastic reduction of its electronic gap when under a conical shape. Furthermore, geometry determines the properties of 2-D materials, and we introduce discrete differential geometry to study them. This geometry arises from particle/atomic positions; it is not based on a parametric continuum, and it applies across broad disciplinary lines.}
\maketitle

\begin{article}
\begin{abstract}
{Lattice {\em Kirigami}, ultra-light metamaterials, poly-disperse aggregates, ceramic nano-lattices, and two-dimensional (2-D) atomic materials share an inherent structural discreteness, and their material properties evolve with their shape. To exemplify the intimate relation among material properties and the local geometry, we explore the properties of phosphorene --a new 2-D atomic material-- in a conical structure, and document a decrease of the semiconducting gap that is directly linked to its non-planar shape. This geometrical effect occurs regardless of phosphorene allotrope considered, and it provides a unique optical vehicle to single out local structural defects on this 2-D material. We also classify other 2-D atomic materials in terms of their crystalline unit cells, and propose means to obtain the local geometry directly from their diverse two-dimensional structures while bypassing common descriptions of shape that are based from a parametric continuum.}
\end{abstract}

\abbreviations{DDG, discrete differential geometry; 2-D, two-dimensional}

\dropcap{T}wo-dimensional (2-D) materials [1-20] 
are discrete surfaces that are embedded on a three-dimensional space. Graphene \cite{gr0,KatsnelsonBook} develops an effective Dirac-like dispersion on the sublattice degree of freedom and other 2-D atomic materials exhibit remarkable plasmonic, polariton, and spin behaviors too [18-20]. 

The properties of 2-D materials are influenced by their local geometry [12-17,21-29], 
making a discussion of the shape of two-dimensional lattices a timely and fundamental endeavor \cite{Maria2010,KamienRMP,Bobenko1,DDG}.
A dedicated discussion of the shape of 2-D materials is given here within the context of nets. Nets are discrete surfaces made from vertices and edges, with vertices given by particle/atomic positions [31-34].
The {\em discrete geometry} that originates from these material nets is richer than its smooth counterpart because the net preserves the structural information of the 2-D lattice completely, yielding exact descriptions of shape that remain accurate as the lattice is subject to arbitrarily large structural deformations \cite{PNASIrvine,Manoharan}, as {\em the particle/atomic lattice {\em becomes} the net itself}.

A discrete geometry helps addressing how strain influences Chemistry \cite{Haddon}, how energy landscapes \cite{WalesBook,Wales1,Wales6} correlate to non-planar shapes, and it provided the basis for a lattice gauge theory for effective Dirac fermions on deformed graphene \cite{usgeo,ACSNano}. In continuing with this program, the optical and electronic properties of phosphorene cones will be linked to their geometry in the present work. At variance with 2-D crystalline soft materials that acquire topological defects while conforming to non-planar shapes [12-15],
the materials considered here have strong chemical bonds that inhibit plastic deformations for strain larger than ten percent \cite{Hone2}.

\begin{figure}[tb]
\centerline{\includegraphics[width=0.45\textwidth]{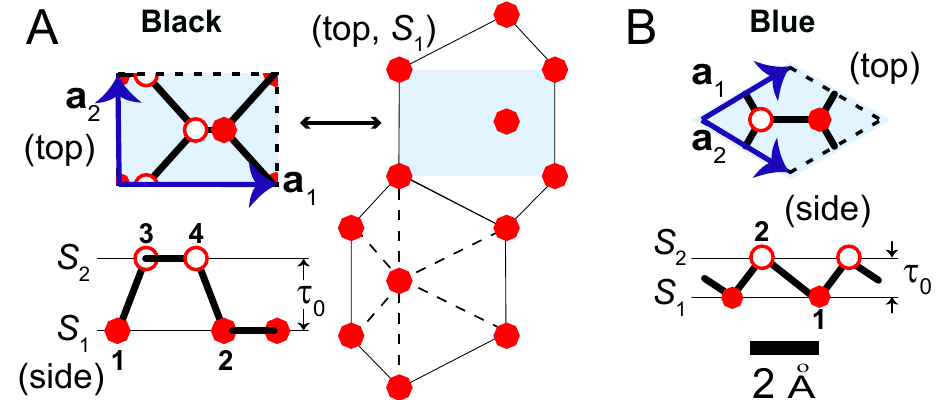}}\label{fig:F1}
\caption{Unit cells for (A) black and (B) blue phosphorene monolayers that are formed by two sublayers ($S_1$ and $S_2$) separated by a distance $\tau_0$. Sublayers in black phosphorene are irregular --distances among nearest atoms belonging to the same sublayer are not identical-- as emphasized by the dashed lines joining atoms belonging to sublayer $S_1$ on the rightmost structure in (A).}
\end{figure}

The study is structured as follows: We build conical structures of black and blue phosphorene, determine their local shape and link this shape to the magnitude of their semiconducting gap. It is clear that a discrete geometry applies for arbitrary 2-D materials.

\begin{figure*}[tb]
\centerline{\includegraphics[width=1.0\textwidth]{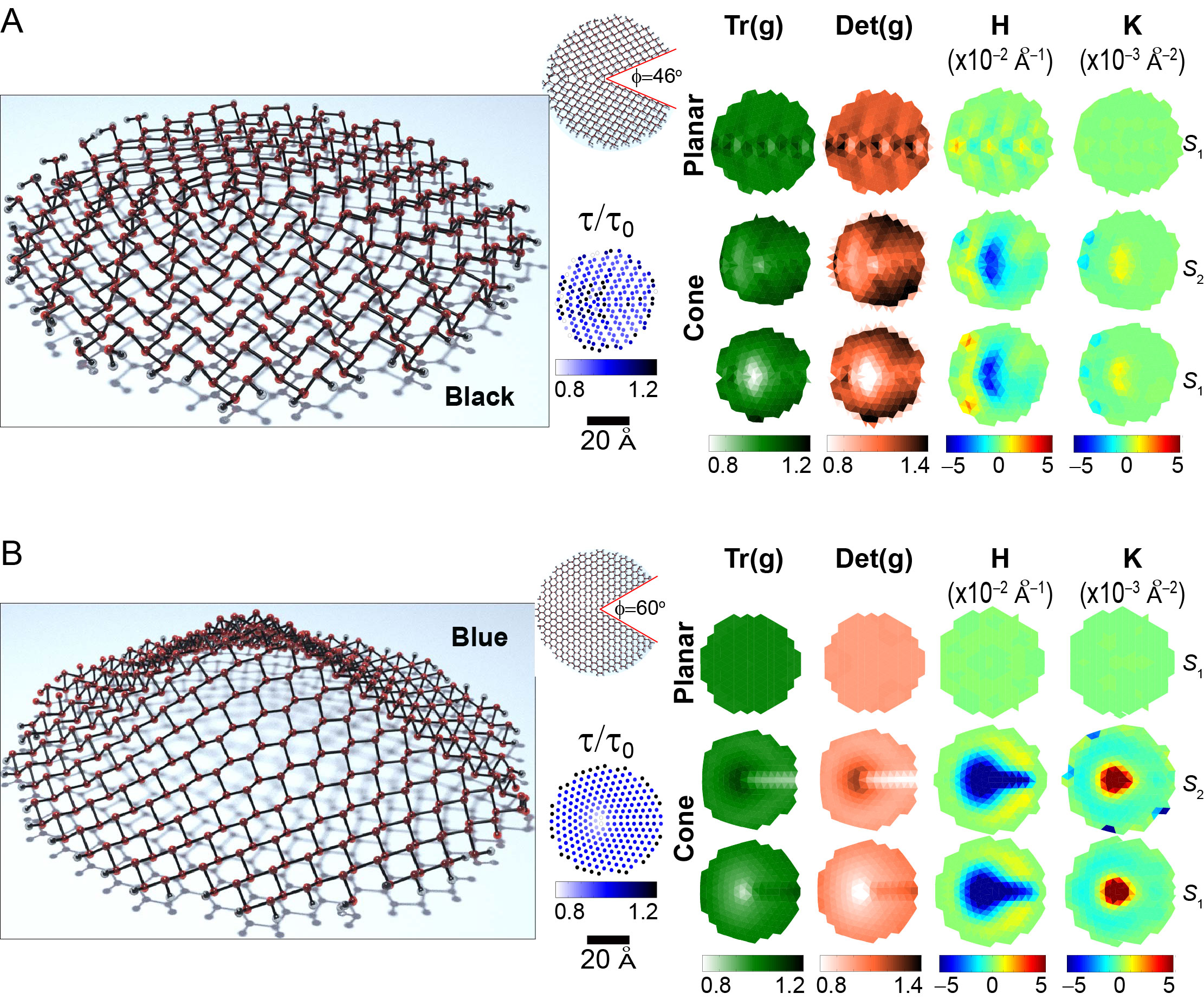}}\label{fig:F1}
\caption{(A) Black and (B) blue phosphorene cones are built by removing the angular segments in white from the planar structures (as illustrated at the upper-right corner of the conical structures), joining the cut structures along the red (disclination) lines, and a subsequent structural optimization. The discrete local geometry of these conical structures at sublayers $S_1$ and $S_2$ is given by four local invariants ($Tr(g)$, $Det(g)$, $H$ and $K$) that are obtained at each atomic position. These invariants are contrasted with the geometry of planar structures  ($Tr(g)=1$, $Det(g)=1$, $H=0$, and $K=0$) that is depicted at the uppermost row for sublayer $S_1$. The change in relative height $\tau/\tau_0$ is shown at each atomic position as well. Features such as: (i) strain induced by the dislocation line in planar black phosphorene; (ii) a sensible compressive strain near the apex; (iii) curvature lacking a radial symmetry; and (iv) the lack of significant changes in $\tau$ on conical black phosphorene, are clearly visible. The disclination line on the blue phosphorene cone is reflected on the metric invariants and on $H$: This disclination has a semi-cylindrical shape, that yields an overall radially-symmetric Gaussian curvature $K$. In addition, a decrease of $\tau$, and an in-plane compression (elongation) at sublayer $S_1$ ($S_2$) near the apex can also be seen on the blue phosphorene cone.}
\end{figure*}

\section{Results and discussion}
Phosphorene \cite{Phosphorene1,Phosphorene2} has many allotropes that are either  semiconducting or metallic depending on their two-dimensional atomistic structure \cite{TomanekBlue,allotropes,Tiling}. The most studied phase, black phosphorene (Fig.~1A), has a semiconducting gap that is tunable with the number of layers, and by in-plane strain \cite{BlackStrain1,BlackStrain2,KatPhos,Peeters}. Theoretical studies of defects on {\em planar} phosphorene indicate that dislocation lines do not induce localized electronic states \cite{Jakobson2}, and algorithms to tile arbitrary planar 2-D phosphorene patterns have been proposed as well \cite{Tiling}. Unit cells of planar monolayers of black and blue phosphorene are displayed on Fig.~1.

Four invariants from the metric ($g$) and curvature ($k$) tensors determine the local geometry of a 2-D manifold \cite{Maria2010,KamienRMP}:
\begin{equation}\label{eq:eq1}
\text{Tr}(g), \/\/ \text{Det}(g),\/\/ H\equiv \text{Tr}(k)/2\text{Tr}(g)\text{, }K\equiv\text{Det}(k)/2\text{Det}(g),
\end{equation}
where Tr (Det) stands for trace (determinant), $H$ is the mean curvature, and $K$ is the Gaussian curvature.

An infinite crystal can be built from these unit cells, and the geometry of such ideal planar structure can be described by $Tr(g)=1$, $Det(g)=1$ (i.e., no strain), $H=0$, and $K=0$ (i.e., no curvature) at both sublayers $S_1$ and $S_2$. In addition, their thickness $\tau$ is equal to $\tau_0$ before a structural distortion sets in: These five numbers quantify the local {\em reference, flat geometry}. (In principle, Tr($g$) and Det($g$) will be functions of the lattice constant, but here they are normalized with respect to their values in the reference crystalline structures seen on Fig.~1, in order to enable direct comparisons of the metric among black and blue phosphorene monolayers. $\tau_0=2.27$  (1.26) \AA{} for black (blue) phosphorene in these planar reference structures.)

Now, structural defects can induce non-zero curvature and strain \cite{Haddon}, and may also be the culprits for the chemical degradation of layered phosphorene. In order to study the consequences of shape on the optical and electronic properties of phosphorene monolayers, we create finite-size conical black and blue phosphorene monolayers, characterize the atomistic geometry, and investigate the influence of shape on their semiconducting gap. (We have proposed that hexagonal boron nitride could slow the degradation process while allowing for a local characterization of phosphorene allotropes \cite{PabloPhosphorene}, and a study of chemical reactivity of non-planar phosphorene will be presented elsewhere.)

Phosphorene cones are built from finite disk-like planar structures that have hydrogen-passivated edges (c.f., Fig.~2). The black phosphorene cone seen in Fig.~\ref{fig:F1}A is created as follows: We remove an angular segment --that subtends a $\phi=46^{\circ}$ angle-- from a planar structure that has a dislocation line \cite{Jakobson2}. The two ``ridges'' that are highlighted by the red segments on the planar structure in Fig.~\ref{fig:F1}A are joined afterwards to create a disclination line. Atoms are placed in positions dictated by an initial (analytical) conical structure, and there is a full structural optimization via molecular dynamics at the {\em ab-initio} level to relieve structural forces throughout (see Methods).

Blue phosphorene has a (buckled) honeycomb structure reminiscent of graphene, so the conical structure seen to the left of Fig.~\ref{fig:F1}B was generated by removing an angular segment subtending a $\phi=60^\circ$ angle on planar blue phosphorene, and following prescriptions similar to those employed in creating the black phosphorene cone afterward.

The subplots displayed to the right on Figs.~2A and 2B show the local discrete geometry at individual atoms (c.f., Eqn.~\eqref{eq:eq1}; see details in the Methods section). For each allotrope, the data is arranged into three rows that indicate the geometry of the planar structure at sublayer $S_1$, and the local geometry of the cones at sublayers $S_1$ and $S_2$. An additional plot shows the value of $\tau/\tau_0$ that tell us of local vertical compression.

There is strain induced by the dislocation line on the planar black phosphorene structure, as indicated by the color variation on the $Tr(g)$ and $Det(g)$ plots, that implies having atoms at closer/longer distances than in the reference structure, Fig.~1. A slight curvature on the black phosphorene planar structure, induced by the dislocation line, is also visible on the $H$ plot in Fig.~2A. The planar blue phosphorene sample does not have any dislocation line, and for that reason the metric shows zero strain ($Tr(g)=1$ and $Det(g)=1$) and zero curvature ($H=0$ and $K=0$) on that reference structure (Fig.~2B).

The black phosphorene conical structure seen on Fig.~2A has the following features: A compression near its apex, as displayed by the white color on the metric invariants; this compression is not radial-symmetric. An asymmetric elongation is visible towards the edges. In addition, there is a radially-asymmetric non-zero curvature; the observed asymmetry reflects the presence of the dislocation/disclination axis, which provides the structure with an enhanced structural rigidity. This rigidity is confirmed by the ratio $\tau/\tau_0$ close to unity, which indicates almost no vertical compression/elongation of this conical structure (Fig.~2A).

The blue phosphorene cone has a more apparent radial symmetry, except for the disclination line that is created by the conical structure, as reflected on the metric invariants and on $H$. There is compression (elongation) at the apex, and elongation (compression) along the disclination line in sublayer $S_1$ ($S_2$). The disclination line has a semi-cylindrical shape and hence a zero radius of curvature along the disclination line, resulting in a zero Gaussian curvature along such line; the Gaussian curvature looks radially-symmetric overall.

Figure~\ref{fig:F1} indicates that the blue phosphorene cone acquires the largest curvatures of these two cones. This is so because the angular segment removed from the planar blue phosphorene sample has a comparatively larger value of $\phi$. One notes the rather smooth curvature at the apex on the blue phosphorene conical structure after the structural optimization.

The change of $\tau$ with respect to $\tau_0$ is created by an out-of-plane strain or by shear. The blue phosphorene cone shows out-of-plane compression near the apex. The distance among planes in black phosphorene is closer to its value in an ideal planar structure throughout, showing scatter around the dislocation line.

The main point from Figure 2 is the strain induced by curvature. The strain pattern observed on that Figure is far more complex than those reported before for planar phosphorene \cite{BlackStrain1,BlackStrain2,KatPhos,Peeters}, and the discrete geometry captures the strain pattern with the precision given by actual atomic positions. We will describe the tools that lead to this geometry based from atoms later on (Methods), but we describe the effect of shape on the material properties of these cones first.

Black and blue phosphorene monolayers are both semiconducting 2-D materials with a direct bandgap, and we investigate how this semiconducting gap evolves with their shape. The semiconducting gap $E_g=$ is equal to 1.1 (2.2) eV for the finite-size planar black (blue) phosphorene monolayers on display in Fig.~2, before the removal of the angular segments. These electronic gaps are highlighted at the rightmost end of Fig.~3. All gaps were computed after a structural optimization, in order to avoid unbalanced forces on these samples that would bias their magnitude (Methods).

We determine size-effects on the semiconducting gap on planar structures first: The magnitude of the gap $E_g$ increases as the number of atoms decreases on strain-free disk-like planar structures; c.f., Fig.~3. For an infinite number of atoms --i.e., for a fully periodic planar 2-D crystal-- the gap converges to the values indicated by the dash and dash-dot lines on Fig.~3; namely, 0.8 eV and 2.0 eV for black and blue phosphorene, respectively. These magnitudes were obtained from standard density-functional theory \cite{Phosphorene1,TomanekBlue}. Although we acknowledge that other methods describe the dielectric properties more accurately on smaller samples \cite{BlackStrain2}, our focus is on the trend the gap follows, and given that {\em the trend is geometrical in nature, it will stand correct} despite of the particular method employed in computing the semiconducting gap.

\begin{figure}[tb]
\centerline{\includegraphics[width=0.45\textwidth]{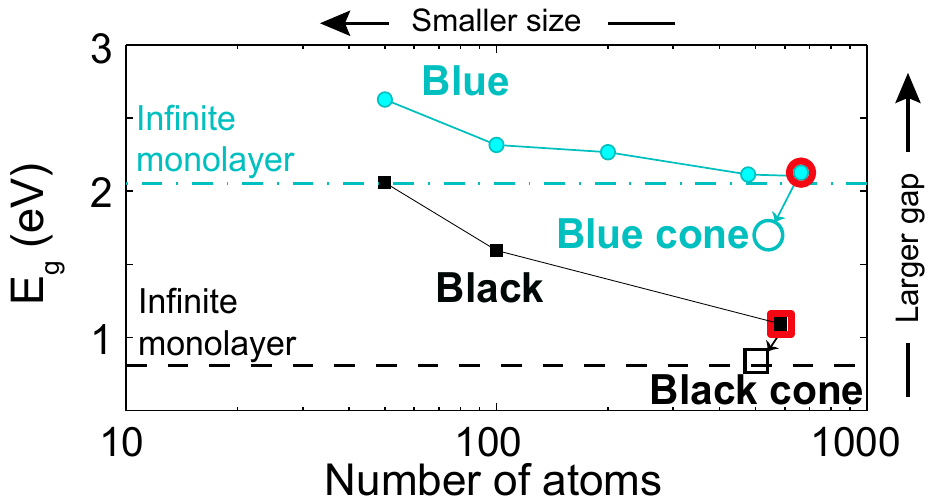}}\label{fig:F3}
\caption{The semiconducting gap of planar black and blue phosphorene monolayer discs increases as the number of atoms decreases. Even though the number of atoms must be reduced to create a conical shape from a planar structure, the gap decreases on conical structures with such decrease of the number of atoms, as indicated by open ticks and the tilted arrows: This reduction of the gap on a conical structure thus has a purely geometrical origin. (Dash and dash-dot lines indicate these gaps when the number of atoms is infinite.)}
\end{figure}

The conical structures have less atoms than their parent planar structures once the angular segments seen in Fig.~2A and 2B are removed. We learned on previous paragraph that the electronic gap increases as the number of atoms is decreased on planar structures. Following that argument, one may expect the semiconducting gap for the conical structures to be larger than the one observed on their parent planar structures.

\begin{figure*}[tb]
\centerline{\includegraphics[width=1.0\textwidth]{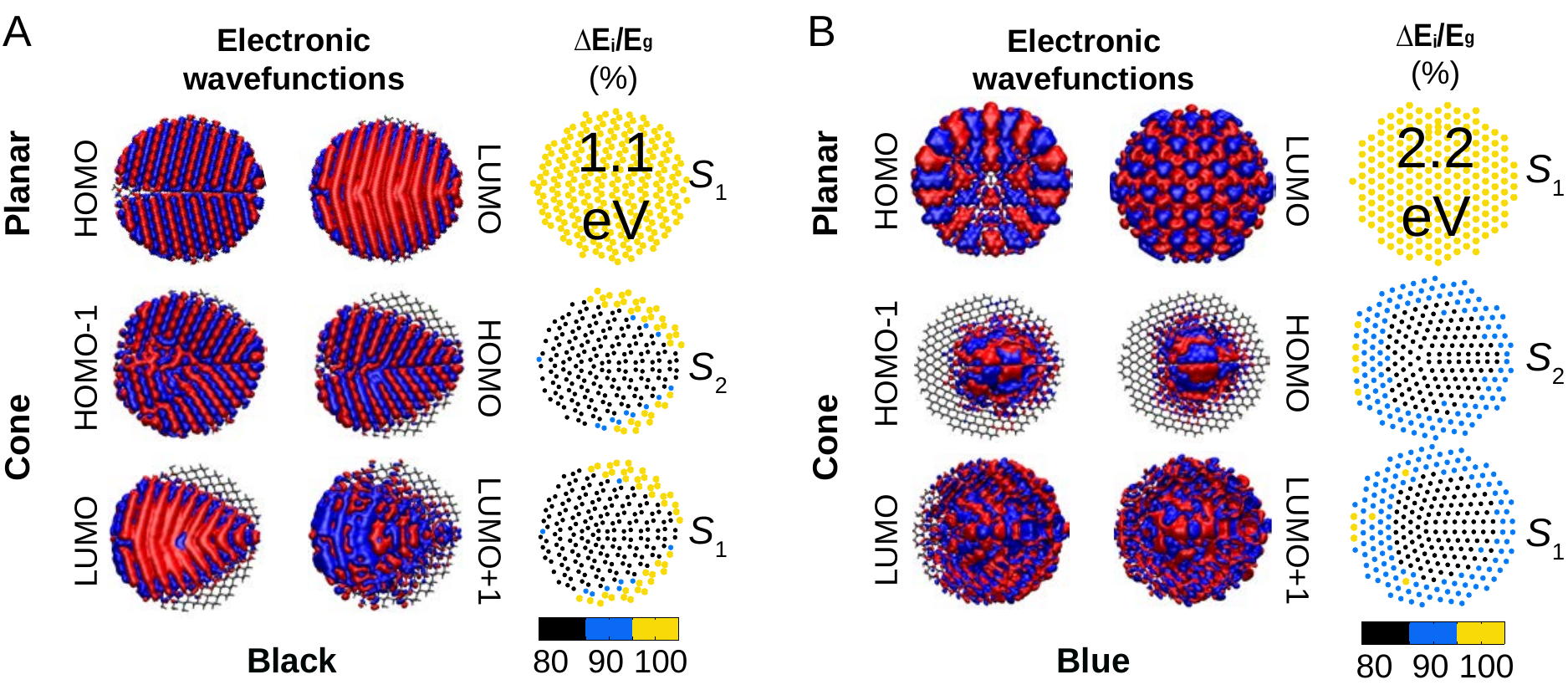}}\label{fig:F4}
\caption{A remarkable reduction of the semiconducting gap occurs due to the strain induced by the conical shape, and regardless of phosphorene allotrope. The local distribution of $\Delta E_i$ correlates with the extent of the electronic wavefunctions shown below and above the Fermi level for (A) black and (B) blue phosphorene.}
\end{figure*}

But instead of increasing with the removal of atoms, the gap decreases dramatically on the conical structures, going down to 0.84 (1.70) eV for the black (blue) phosphorene cone, as indicated by the open red symbols, and the tilted arrows in Fig.~3: {\em Shape influences phosphorene's material properties.} A practical consequence of this result is that optical probes could provide a measure of the local shape of phosphorene. The relation among the semiconducting gap and a non-planar shape has been observed in transition metal dichalcogenide monolayers recently \cite{strainGuinea2013}, and we establish here that structural defects on semiconducting 2-D materials create a similar effect. Let us next address the reason for such reduction of the semiconducting gap.

The semiconducting gap is a global material property. On finite samples it is possible to project the electronic density onto individual atoms to learn about the spatial localization of the electronic states below and above the Fermi level, whose energy difference is equal to the electronic gap. Those states are known as the highest occupied molecular orbital (HOMO) and the lowest unoccupied molecular orbital (LUMO), and they were spin-degenerate in all the samples we studied. Additionally, the $n-th$ electronic orbital below (above) the HOMO (LUMO) is labeled HOMO$-n$ (LUMO$+n$).

The HOMO and LUMO states cover the entire planar black and blue phosphorene structures (c.f., Fig.~4) and are thus {\em delocalized}. On the other hand, the conical samples have orbitals below and above the Fermi energy that display a certain amount of localization.

Such effect is most visible for the HOMO and HOMO$-1$ states for the blue phosphorene cone in Fig.~4B. Thus, unlike a dislocation line \cite{Jakobson2} the pentagon defect that is responsible for the curvature of the blue conical structure is localizing the HOMO state within a $\sim10$ \AA{} radius. Given that {\em the pentagon defect originates curvature itself, we conclude that curvature leads to a reduction of the electronic gap} on this system.

Given the localization observed on some of these orbitals, we can further ask: At a given atomic position, what is the first orbital with a non-zero density at such location below and above the Fermi energy? This question can be rephrased in terms of the difference in energy among the first orbitals below and above the Fermi level that have a non-zero probability density at a given atomic location: On the third column on Figs.~4A and 4B we display such energy difference $\Delta E_i$ at every phosphorus atom $i$. There exists a clear correlation among $\Delta E_i$ and the localization pattern of the orbitals below and above the Fermi energy, as it should be the case. The reduction on the semiconducting gap on the cones is emphasized by normalizing $\Delta E_i$ in terms of $E_g$, which yields an twenty-percent reduction of the semiconducting gap for both allotropes: A non-planar geometry that is created by structural defects on any phosphorene allotrope will lead to a sizeable reduction of the semiconducting gap.

To end this work, we must state that the discrete geometry used to tell the shape of the phosphorene conical structures applies to other 2-D atomic materials with varied unit cells; some of which are listed below:
\begin{enumerate}
\item{} Regular honeycomb lattices (graphene \cite{KatsnelsonBook,usgeo,ACSNano,TomanekCurvature} and hexagonal boron nitride (hBN) \cite{BNsinglelayer}.
\item{} Low-buckled honeycomb hexagonal lattices (silicene and germanene \cite{silicene}. We established that freestanding stanene \cite{stanene2} is not the structural ground state.
\item{} ``High-buckled'' hexagonal close-packed bilayers of bismuth \cite{Palacios}, tin and lead \cite{ussubmitted}.
\item{}  Thin trigonal-prismatic transition metal dichalcogenides \cite{MX2,reviewMX2}.
\item{} Materials with buckled square unit cells --{\em quad-graphs}~\cite{Bobenko1}-- such as AlP \cite{Hennig1}.
\end{enumerate}
Low-buckled structures (silicene, germanene, blue phosphorus), hexagonal close-packed bilayers (bismuth, tin and lead) and black phosphorus have two parallel sublayers $S_1$ and $S_2$. Transition-metal dichalcogenides and AlP have three parallel sublayers $S_1-S_3$. Structures 1$-$4 are equilateral triangular lattices with a basis, for which individual planes represent regular equilateral triangular nets; and structure 5 realizes a regular quad-graph. The discrete geometry clearly stands for other 2-D materials that do not form strong directional bonds \cite{Science2003,Nature2010Irvine} as well.

\section{Concluding remarks}

Shape is a fundamental handle to tune the properties of 2-D materials, and the discrete geometry provides the most accurate description of two-dimensional material nets. This geometry was showcased on non-planar phosphorene allotropes for which the electronic gap decreases by twenty-percent with respect to its value on planar structures. The discrete geometry can be thus be used to correlate large structural deformations to intended functionalities on 2-D materials with arbitrary shapes.

\section{Methods}
\subsection{Creation of conical structures} Consider structures having atomic positions at planar discs; these positions can be parameterized by $r_i$, $\theta_i$, and $z_i=\{0,\tau_0\}$. Calling $\phi$ the angular segments being removed from these planar structures, the range of the angular variable $\theta_i$ is $[0,2\pi-\phi]$, and the conical structures are initially built by the following transformation: $r_i'=r_i\sqrt{1-\frac{\phi^2}{4\pi^2}}$, $\theta_i'=\frac{2\pi}{2\pi-\phi}\theta_i$, and $z_i'=z_i-r_i\frac{\phi}{2\pi}$. These conical structures containing about five hundred atoms undergo a structural optimization via {\em ab-initio} (Car-Parinello) molecular dynamics \cite{CP} with the {\em SIESTA} code \cite{SIESTA,PBE}, until forces are smaller than 0.04 eV/\AA.

\subsection{Calculation of the electronic gap of phosphorene samples}
We obtain $\Delta E_i$ as follows: Let $E_F$ be the charge neutrality level or Fermi energy, and $\rho_i(E)$ the density of electronic states projected onto atom $i$. We call $E^A_i$ ($E^B_i$) the first energy level observed on $\rho_i(E)$ lying above (below) $E_F$ at atom $i$, and report $\Delta E_i/E_g\equiv (E^A_i-E^B_i)/E_g$ on Fig.~4. We note that all structures we worked with had a final net zero spin polarization on spin-polarized calculations.

\subsection{A discrete geometry based on triangulations at atomic positions}
There is an {\em extrinsic} and continuum (Euclidean) geometry in which material objects exist. These material objects are made out of atoms that take specific locations to generate their own {\em intrinsic} shape.

\begin{figure}[h]
\centerline{\includegraphics[width=0.45\textwidth]{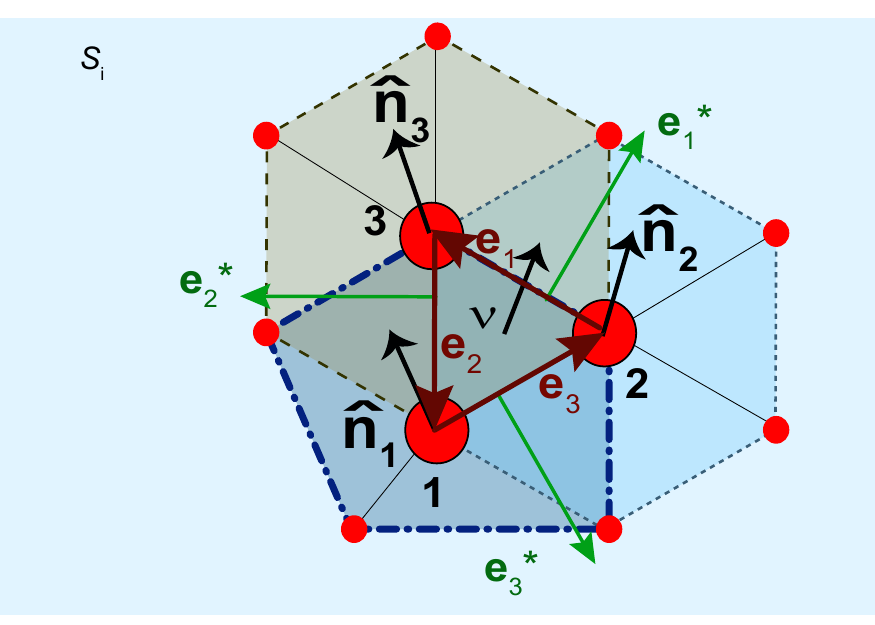}}\label{fig:F5}
\caption{Discrete tensors based on triangulations are expressed in terms of averaged normals $\hat{\mathbf{n}}_j$, edges $\mathbf{e}_j$, the normal of the triangle $\boldsymbol{\nu}$, and dual edges $\mathbf{e}_j^*\equiv\mathbf{e}_j\times \boldsymbol{\nu}$ ($j=1,2,3$). These tensors can be employed in quad-nets, irregular lattices, and structures containing defects as highlighted by the pentagon surrounding atom $\mathbf{1}$.}
\end{figure}

The intrinsic shape of 2-D atomic materials can be idealized as a continuum: The basic assumption of continuum mechanics is that a continuum shape is justified at length scales $l$ much larger than interatomic distances that are characterized by a lattice parameter $a_0$ ($l>>a_0$). The common understanding of shape arises from within this continuum perspective that is based on the differential geometry of two-dimensional manifolds. This continuum approximation is valid down to interatomic scales ($l\leq a_0$) for slowly-varying deformations, but not when curvature concentrates at atomic-scale bonds or pleats, such as in the examples provided in Ref.~\cite{kirigami} and \cite{Nature2010Irvine}. And so, while these sharp structures may be disregarded or approximated within the context of a continuum surface, we hold the opinion that an intrinsic geometry that is exact at the atomic scale must form part of the theoretical tool-set to deal with 2-D materials. The discrete geometry to be described in the next paragraphs bypasses descriptions of shape that are based on an effective continuum, and it brings an understanding of the shape of material nets at a fundamental level.

Consider three {\em directed} edges $\mathbf{e}_1$, $\mathbf{e}_2$ and $\mathbf{e}_3$, such that $\mathbf{e}_1+\mathbf{e}_2+\mathbf{e}_3=\mathbf{0}$, and define $Q_l^I\equiv\mathbf{e}_l\cdot\mathbf{e}_l$ ($l=1,2,3$), representing the square of the smallest {\em finite}  distance among atoms on the 2-D lattice. This is the discrete analog of the infinitesimal length element $ds^2$ (c.f., Fig.~5).

Now consider the change in orientation among normals $\hat{\mathbf{n}}_j$ and $\hat{\mathbf{n}}_k$, and project such variation onto their common edge $\mathbf{e}_l$: This is is, define $Q_l^{II}\equiv (\hat{\mathbf{n}}_k-\hat{\mathbf{n}}_j)\cdot \mathbf{e}_l$ (see Fig.~5; $j,k,l$ are permutations of integers 1, 2, and 3). The reader may recall that the curvature tensor is defined as $\hat{n}\cdot\frac{\partial g_{\alpha}}{\partial \xi_{\beta}}$, but the discrete tensor carries a constant edge, and it rests on changes of the local normals instead. In previous Equation, $\hat{\mathbf{n}}_l$ is the average over individual normals at triangulated area elements within the polygon surrounding atom $l$ and highlighted by dashed lines. The dual edge is defined by $\mathbf{e}_l^*\equiv \mathbf{e}_l\times \boldsymbol{\nu}$, with $\boldsymbol{\nu}$ the normal to the triangle formed by atoms $\mathbf{1}$, $\mathbf{2}$ and $\mathbf{3}$ and $A_T$ is the triangle area ($-\mathbf{e}_2\times \mathbf{e}_1=2A_T\boldsymbol{\nu}$).

This way, the discrete metric tensor takes the following form on the basis given by dual edges \cite{Alejandro}:
\begin{equation}\label{eq:eq8}
g=-\frac{1}{8A_0^2}\sum_{(j,k,l)}(Q_j^I - Q_k^I- Q_l^I)\mathbf{e}_j^*\otimes\mathbf{e}_j^*,
\end{equation}
with $A_0$ the area of the triangulated area element at the reference (non-deformed and defect-free) plane.

The discrete curvature tensor has an identical structure:
\begin{equation}\label{eq:eq7}
k=-\frac{1}{8A_T^2}\sum_{(j,k,l)}(Q_j^{II}-Q_k^{II}-Q_l^{II})\mathbf{e}_j^*\otimes\mathbf{e}_j^*.
\end{equation}
The parenthesis $(j,k,l)$ indicates a sum of three terms, as follows: ($j=1$, $k=2$, $l=3$), (2, 3, 1), and (3, 1, 2). Eqns.~\eqref{eq:eq8} and \eqref{eq:eq7}  become 3$\times$3 matrices with explicit values for $Q_l^I$, $Q_l^{II}$ and $\mathbf{e}_j^*$ from atomic positions (Fig.~5). For instance, the discrete curvature tensor has eigenvalues \{0, $k_1$, $k_2$\} at each triangulated area element, yielding $H=(k_1+k_2)/2$ and $K=k_1k_2$. The geometrical invariants reported at point $\mathbf{j}$ are averages over their values at individual triangles sharing this vertex.

Topological defects (exemplified by a pentagon seen by a dot-dash line on Fig.~5) break translation symmetry, making it impossible to recover a crystalline structure by means of elastic deformations. Yet,
Eqns.~\eqref{eq:eq8} and \eqref{eq:eq7} provide a geometry along topological defects seamlessly still, as seen in the geometry provided in Fig.~2.

The computation of the distribution of the thickness $\tau$ in both conical structures was performed by finding for each atom in the lower sublayer $S_1$ the three nearest neighbors on the upper sublayer $S_2$. The distance between the centroid these three atoms and the atom in question at the $S_1$ sublayer amounts to the thickness $\tau$ of the structure at that specific atom.

\begin{acknowledgments}
This work was supported by the Arkansas Biosciences Institute (S.B.L.) and by NSF (H.T., EFRI-1433311). Calculations were performed at Arkansas' {\em Razor II} and TACC's {\em Stampede} (NSF-XSEDE, TG-PHY090002).
\end{acknowledgments}

\noindent{\small Author contributions: A.P.S.J., H.T., E.O.H. and S.B.L. designed research. all authors performed research, analyzed data and wrote the paper.}
\noindent{\small The authors declare no conflict of interest.}

\end{article}

\begin{thebibliography}{10}
\bibitem{gr0}
Novoselov KS, Geim AK, Morozov SV, Jiang D, Katsnelson MI, Grigorieva IV,
  Dubonos SV, Firsov AA
\newblock (2005) Two-dimensional gas of massless Dirac fermions in graphene. {\em Nature} {\bf 438}, 197--200.

\bibitem{KatsnelsonBook}
Katsnelson MI
\newblock (2012) {\em Graphene: Carbon in Two Dimensions}.
\newblock (Cambridge U. Press), first edition.

\bibitem{NovoPNAS}
Novoselov KS, Jiang D, Schedin F, Booth TJ, Khotkevich VV, Morozov SV, Geim AK
\newblock (2005) Two-dimensional atomic crystals. {\em Proc. Natl. Acad. Sci. (USA)} {\bf 102}, 10451.

\bibitem{BNsinglelayer}
Jin, C, Lin, F, Suenaga, K, Iijima, S
\newblock (2009) Fabrication of a freestanding boron nitride single layer and its defect assignments. {\em Phys. Rev. Lett.} {\bf 102}, 195505.

\bibitem{silicene}
Cahangirov S, Topsakal M, Akt{\"u}rk E, Sahin H, Ciraci S
\newblock (2009) Two- and one-dimensional honeycomb structures of silicon and germanium. {\em Phys. Rev. Lett.} {\bf 102}, 236804.

\bibitem{Bi2Se32D}
Zhang Y, He K, Chang C-Z, Song C-L, Wang L-L, Chen X, Jia J-F, Fang Z,  Dai X, Shan W-Y, Shen S-Q, Niu Q, Qi X-L, Zhang S-C, Ma X-C, Xue Q-K
\newblock (2010) Crossover of the three-dimensional topological insulator Bi$_2$Se$_3$ to the two-dimensional limit. {\em Nature Phys.} {\bf 6}, 584--588.

\bibitem{Hennig1}
Zhuang HL, Singh AK, Hennig RG
\newblock (2013) Computational discovery of single-layer III-V materials. {\em Phys. Rev. B} {\bf 87}, 165415.

\bibitem{Phosphorene1}
Liu H, Neal AT, Zhu Z, Luo Z, Xu X, Tom{\'a}nek D, Ye PD
\newblock (2014) Phosphorene: An unexplored 2D semiconductor with a high hole mobility. {\em ACS Nano} {\bf 8}, 4033–-4041.

\bibitem{Phosphorene2}
Li L, Yu Y, Ye GJ, Ge Q, Ou X, Wu H, Feng D, Chen X, Zhang Y
\newblock (2014) Black phosphorus field-effect transistors. {\em Nature Nanotech.} {\bf 9}, 372--377.

\bibitem{KaneMele}
Kane CL, Mele EJ
\newblock (2005) Quantum spin Hall effect in graphene. {\em Phys. Rev. Lett.} {\bf 95}, 226801.

\bibitem{kirigami}
Castle T, Cho Y, Gong X, Jung E, Sussman DM, Yang S, Kamien RD
\newblock (2014) Making the cut: Lattice {\em kirigami} rules. {\em Phys. Rev. Lett.} {\bf 113}, 245502.

\bibitem{Science2003}
Bausch AR, Bowick MJ, Cacciuto A, Dinsmore AD, Hsu MF, Nelson DR, Nikolaides MG, Travesset A, Weitz DA
\newblock (2003) Grain boundary scars and spherical crystallography. {\em Science} {\bf 299}, 1716--1718.

\bibitem{PNAS2006}
Vitelli V, Lucks JB, Nelson DR
\newblock (2006) Crystallography on curved surfaces. {\em Proc. Natl. Acad. Sci. (USA)} {\bf 103}, 12323--12328.

\bibitem{Nature2010Irvine}
Irvine WTM, Vitelli V, Chaikin PM
\newblock (2010) Pleats in crystals on curved surfaces. {\em Nature} {\bf 468}, 947--951.

\bibitem{PNASIrvine}
Irvine WTM, Hollingsworth AD, Grier DG, Chaikin PM
newblock (2013) Dislocation reactions, grain boundaries, and irreversibility in two-dimensional lattices using topological tweezers.
{\em Proc. Natl. Acad. Sci. (USA)} {\bf 110}, 15544--15548.

\bibitem{Manoharan}
Gomes KK, Mar W, Ko W, Guinea F, Manoharan HC
\newblock (2012) Designer Dirac fermions and topological phases in molecular graphene.
{\em Nature} {\bf 483}, 306--310.

\bibitem{PNASGrason}
Grason GM, Davidovitch B
\newblock (2013) Universal collapse of stress and wrinkle-to-scar transition in spherically confined crystalline sheets.
{\em Proc. Natl. Acad. Sci. (USA)} {\bf 110}, 12893--12898.

\bibitem{TerronesWS2}
Guti{\'e}rrez HR, Perea-Lopez, N, Elias AL, Berkdemir A, Wang B, Lv R,
  L{\'o}pez-Ur{\'\i}as F, Crespi VH, Terrones H, Terrones M
\newblock (2013) Extraordinary room-temperature photoluminescence in triangular WS$_2$ monolayers. {\em Nano Lett.} {\bf 13}, 3447--3454.

\bibitem{Science2014}
Dai S, Fei Z, Ma Q, Rodin AS, Wagner M, McLeod AS, Liu MK, Gannett W,
  Regan W, Watanabe K, Taniguchi T, Thiemens M, Dominguez G, Castro Neto AH,
  Zettl A, Keilmann F, Jarillo-Herrero P, Fogler MM, Basov DN
\newblock (2014) Tunable phonon polaritons in atomically thin van der Waals crystals of boron nitride. {\em Science} {\bf 343}, 1125--1129.

\bibitem{valleytronicsMoS2}
Zibouche N, Philipsen P, Kuc A, Heine T
\newblock (2014) Transition-metal dichalcogenide bilayers: Switching materials for spintronic and valleytronic applications. {\em Phys. Rev. B} {\bf 90}, 125440.

\bibitem{MonicaPNAS2013}
Funkhouser CM, Sknepnek R, Shimi T, Goldman AE, Goldman RD, {Olvera~de~la~Cruz} M
\newblock (2013) Mechanical model of blebbing in nuclear lamin meshworks. {\em Proc. Natl. Acad. Sci. (USA)} {\bf 110}, 3248.

\bibitem{monica2011}
Vernizzi G, Sknepnek R, {Olvera~de~la~Cruz} M
\newblock (2011) Platonic and Archimedean geometries in multicomponent elastic membranes. {\em Proc. Natl. Acad. Sci. (USA)} {\bf 108}, 4292--4296.

\bibitem{monica2014}
Yao Z, {Olvera~de~la~Cruz} M
\newblock (2014) Polydispersity-driven topological defects as order-restoring excitations. {\em Proc. Natl. Acad. Sci. (USA)} {\bf 111}, 5094--5099.

\bibitem{Maria2010}
Vozmediano MAH, Katsnelson MI, Guinea F
\newblock (2010) Gauge fields in graphene. {\em Phys. Rep.} {\bf 496}, 109--150.

\bibitem{strainGuinea2013}
Castellanos-Gomez A, Rold{\'a}n R, Capelluti E, Buschema M, Guinea F, {van der Zant} HSJ, Steele GA
\newblock (2013) Local strain engineering in atomically thin MoS$_2$. {\em Nano Lett.} {\bf 13}, 5361--5366.

\bibitem{BlackStrain1}
Rodin AS, Carvalho A, Castro Neto AH
\newblock (2014) Strain-induced gap modification in black phosphorus. {\em Phys. Rev. Lett.} {\bf 112}, 176801.

\bibitem{BlackStrain2}
Fei R, Yang L
\newblock (2014) Strain-engineering the anisotropic electrical conductance of few-layer black phosphorus. {\em Nano Lett.} {\bf 14}, 2884--2889.

\bibitem{inspired}
van der Zande A, Hone J
\newblock (2012) Inspired by strain. {\em Nature Photonics.} {\bf 6}, 804--806.

\bibitem{nanolattices}
Meza LR, Das S, Greer JR
\newblock (2014) Strong, lightweight, and recoverable three-dimensional ceramic nanolattices. {\em Science} {\bf 345}, 1322--1326.

\bibitem{KamienRMP}
Kamien RD
\newblock (2002) The geometry of soft materials: a primer.
{\em Rev. Mod. Phys.} {\bf 74}, 953--971.

\bibitem{Bobenko1}
Bobenko AI, Suris YB
\newblock (2008) {\em Discrete Differential Geometry. Integrable Structure},
  Graduate Studies in Mathematics.
\newblock (AMS, Providence, RI) Vol.{}~98, first
  edition.

\bibitem{DDG}
Bobenko AI, Schr{\"o}der P, Sullivan JM, Ziegler GM. (2008)
{\em Discrete differential geometry}, Oberwolfach Seminars,
\newblock (Birkh{\"a}user), 1st edition, pp. 175--188.

\bibitem{SIGGRAPH}
Grinspun E, Wardetzky M, Desbrun M, Schr{\"o}der P, eds.
\newblock (2008) {\em Discrete Differential Geometry: An Applied Introduction}.
\newblock Available at: http://ddg.cs.columbia.edu/.

\bibitem{chinese}
Xu Z, Xu G
\newblock (2009) Discrete schemes for Gaussian curvature and their convergence. {\em Comp. Math. Appl.} {\bf 57}, 1187--1195.

\bibitem{Haddon}
Haddon R
\newblock (1993) Chemistry of the fullerenes: The manifestation of strain in a class of continuous aromatic molecules. {\em Science} {\bf 261}, 1545--1550.

\bibitem{WalesBook}
Wales DJ
\newblock (2004) {\em Energy Landscapes: Applications to Clusters, Biomolecules
  and Glasses}.
\newblock (Cambdridge U. Press, Cambridge), first edition.

\bibitem{Wales1}
Wales DJ, Salamon P
\newblock (2014) Observation time scale, free-energy landscapes, and molecular symmetry. {\em Proc. Natl. Acad. Sci. (USA)} {\bf 111}, 617--622.

\bibitem{Wales6}
Kusumaatmaja H , Wales DJ
\newblock (2013) Defect motifs for constant mean curvature surfaces. {\em Phys. Rev. Lett.} {\bf 110}, 165502.

\bibitem{usgeo}
Pacheco-Sanjuan AA, Wang Z, Pour-Imani H, Vanevi{\'c} M, Barraza-Lopez S
\newblock (2014) Graphene's morphology and electronic properties from discrete differential geometry. {\em Phys. Rev. B} {\bf 89}, 121403(R).

\bibitem{ACSNano}
Pacheco Sanjuan AA, Mehboudi M, Harriss EO, Terrones H, Barraza-Lopez S
\newblock (2014) Quantitative chemistry and the discrete geometry of conformal atom-thin crystals. {\em ACS Nano} {\bf 8}, 1136--1146.

\bibitem{Hone2}
Lee G-H, Cooper RC, An SJ, Lee S, {van der Zande} A, Petrone N, Hammerberg AG, Lee C, Crawford B, Oliver W, Kysar JW, Hone J
\newblock (2013) High-strength chemical-vapor–deposited graphene and grain boundaries. {\em Science} {\bf 340}, 1073--1076.

\bibitem{TomanekBlue}
Zhu Z, Tom{\'a}nek D
\newblock (2014) Semiconducting layered blue phosphorus: A computational study. {\em Phys. Rev. Lett.} {\bf 112}, 176802.

\bibitem{allotropes}
Guan J, Zhu Z, Tom{\'a}nek D
\newblock (2014) Phase coexistence and metal-insulator transition in few-layer phosphorene: A computational study. {\em Phys. Rev. Lett.} {\bf 113}, 046804.

\bibitem{Tiling}
Guan J, Zhu Z, Tom\'anek D
\newblock (2014) Tiling phosphorene. {\em ACS Nano} {\bf 8} 12763--12768.


\bibitem{KatPhos}
Rudenko AN, Katsnelson MI
\newblock (2014) Quasiparticle band structure and tight-binding model for single- and bilayer black phosphorus. {\em Phys. Rev. B} {\bf 89} 201408(R).

\bibitem{Peeters}
Cakir D, Sahin H, Peeters FM
\newblock (2014) Tuning of the electronic and optical properties of single-layer black phosphorus by strain. {\em Phys. Rev. B} {\bf 90} 205421.

\bibitem{Jakobson2}
Liu Y, Xu F, Zhang Z, Penev ES, Yakobson BI
\newblock (2014) Two-dimensional mono-elemental semiconductor with electronically inactive defects: The case of phosphorus. {\em Nano Lett.} DOI: 10.1021/nl5021393.

\bibitem{PabloPhosphorene}
Rivero P, Horvath CM, Zhu Z, Guan J, Tom{\'a}nek D, Barraza-Lopez S
\newblock (2015) Simulated scanning tunneling microscopy images of few-layer phosphorus capped by graphene and hexagonal boron nitride monolayers. {\em Phys. Rev. B} {\bf 91}, 115413.





\bibitem{TomanekCurvature}
Guan J, Jin Z, Zhu Z, Chuang C, Jin B-Y, Tom{\'a}nek D
\newblock (2014) Local curvature and stability of two-dimensional systems. {\em Phys. Rev. B} {\bf 90}, 245403.

\bibitem{stanene2}
Xu Y, Yan B, Zhang H-J, Wang J, Xu G, Tang P, Duan W, Zhang S-C
\newblock (2013) Large-gap quantum spin Hall insulators in tin films. {\em Phys. Rev. Lett.} {\bf 111}, 136804.

\bibitem{Palacios}
Sabater C, Gos{\'a}lbez-Mart{\'\i}nez D, Fern{\'a}ndez-Rossier J, Rodrigo
  JG, Untiedt C, Palacios JJ
\newblock (2013) Topologically protected quantum transport in locally exfoliated bismuth at room temperature. {\em Phys. Rev. Lett.} {\bf 110}, 176802.

\bibitem{ussubmitted}
Rivero P, Yan J-A, Garc{\'\i}a-Su{\'a}rez VM, Ferrer J, Barraza-Lopez, S
\newblock (2014) Stability and properties of high-buckled two-dimensional tin and lead. {\em Phys. Rev. B} {\bf 90}, 241408(R).

\bibitem{MX2}
Wang QH, Kalantar-Zadeh K, Kis A, Coleman JN, Strano MS
\newblock (2012) Electronics and optoelectronics of two-dimensional transition metal dichalcogenides. {\em Nature Nanotech.} {\bf 7}, 699--712.

\bibitem{reviewMX2}
Chhowalla M, Shin HS, Eda G, Li L-J, Loh KP, Zhang H
\newblock (2013) The chemistry of two-dimensional layered transition metal dichalcogenide nanosheets. {\em Nature Chem.} {\bf 5}, 263--275.

\bibitem{CP}
Car R, Parinello M
\newblock (1985) Unified approach for molecular dynamics and density-functional theory. {\em Phys. Rev. Lett.} {\bf 55}, 2471.

\bibitem{SIESTA}
Soler J, Artacho E, Gale J, Garc{\'\i}a A, Junquera J, Ordej{\'o}n P, S{\'a}nchez-Portal D
\newblock (2002) The SIESTA method for ab-initio order-N materials simulation. {\em J. Phys.: Condens. Matter} {\bf 14}, 2745--2779.

\bibitem{PBE}
Perdew JP, Burke K, Ernzerhof M
\newblock (1996) Generalized gradient approximation made simple. {\em Phys. Rev. Lett.} {\bf 78}, 3865.


\bibitem{Alejandro}
Weischedel C, Tuganov A, Hermansson T, Linn J, Wardetzky M
\newblock (2012) Construction of discrete shell models by geometric finite differences. In {\em The 2nd joint conference on multibody system dynamics}. Stutgart, Germany.



\end{thebibliography}
\end{document}